\documentclass[11pt]{article}
\usepackage{natbib,fullpage,amssymb,stmaryrd,hyperref,graphicx}

\pdfoutput=1

\newcommand{\iid}{\stackrel{\rm iid}{\sim}}
\newcommand{\etr}{ {\rm etr}}
\newcommand{\tr}{{\rm tr}}
\renewcommand{\a}[1]{\mbox{\bf{#1}} }
\newcommand{\m}[1]{\mbox{\boldmath{${\rm #1}$}}}
\renewcommand{\v}[1]{\mbox{\boldmath{${\rm #1}$}}}

\newcommand{\Exp}[1]{{\rm E}[ \ensuremath{ #1 } ]  }
\newcommand{\Var}[1]{{\rm Var}[ \ensuremath{ #1 } ]  }
\newcommand{\Cov}[1]{{\rm Cov}[ \ensuremath{ #1 } ]  }
\newcommand{\Sig}{\mathbf\Sigma}

\newcommand{\Om}{\mathbf\Omega}
\newcommand{\Ps}{\mathbf\Psi}

\newtheorem{prop}{Proposition}[section]

\begin{document}

\title{Separable covariance arrays via the Tucker product, with applications to multivariate relational data} 
\author{Peter D. Hoff \thanks{Departments of Statistics and Biostatistics ,
University of Washington,
Seattle, Washington 98195-4322.
Web: \href{http://www.stat.washington.edu/hoff/}{\tt http://www.stat.washington.edu/hoff/}. } }
\date{\today} 
\maketitle

\begin{abstract}
Modern datasets are often in the form of matrices 
or arrays, 
potentially having correlations along each set of data indices. 
For example, data involving 
repeated measurements of several variables over time may 
exhibit temporal correlation as well as  correlation among the variables. 
A possible model for matrix-valued data is the 
class of matrix normal distributions, 
which is parametrized by two covariance matrices, one for each index set of the 
data.  
In this article we describe an extension of the matrix normal model 
to accommodate multidimensional data arrays, or tensors. 
We generate a class of array normal distributions 
by applying a group of multilinear transformations to an array 
of independent standard normal random variables. 
The covariance structures of the resulting class 
take the form of outer products of dimension-specific covariance matrices. 
We derive some properties of these covariance structures and the 
corresponding array normal distributions, discuss 
maximum likelihood 
and  Bayesian estimation 
of covariance parameters 
and 
illustrate the model in an analysis of multivariate longitudinal network data. 
\end{abstract}

\noindent {\it Some key words}: Gaussian, matrix normal, multiway data, network, tensor, Tucker decomposition.

\section{Introduction} 
This article provides a construction of and estimation for a class of covariance models and  Gaussian probability 
distributions for array data
consisting of multi-indexed values $\a Y = \{  y_{i_1},\ldots, y_{i_K} : 
  i_k\in\{ 1,\ldots, m_k\},k=1,\ldots, K\}$.  
Such data have become common in many scientific disciplines, including the social and biological sciences. Researchers often gather relational data measured on pairs of units, where the population of units may consist of 
people, genes, websites or some other set of objects.  Data on a single relational 
variable  is often represented by a ``sociomatrix''
 $\m Y = \{ y_{i,j},i\in \{1,\ldots, m\}, j\in \{1,\ldots, m\} , i\neq j\}$, 
a square matrix with an undefined diagonal, where $y_{i,j}$ represents the 
relationship between nodes $i$ and $j$. 

Multivariate  relational data include 
multiple relational measurements on the same node set, 
possibly gathered under different conditions 
or at different time points. Such data can be represented as a multiway array. 
For example, in this article we will analyze  data on trade of 
several commodity classes 
between a set of countries over several years.  
These data can be represented as a four-way array $\a Y = \{ y_{i,j,k,t} \}$, 
where $y_{i,j,k,t}$ records the volume of exports of commodity $k$ from country $i$ to country $j$ in year $t$. 
For such data it is often of interest  to identify similarities or 
correlations among data corresponding to the objects of a given index set. 
For example, one may want to identify 
nodes of a network that behave similarly 
across levels of the other factors of the  array. 
For temporal datasets it may be important to describe correlations 
among data from adjacent time points.  In general, it may be desirable 
to estimate or account for dependencies along each index set of the array. 

For 
matrix valued data, 
such 
considerations have led to the use of separable covariance estimates, 
whereby  the covariance of a population of matrices is estimated as being
 $\Cov{ {\rm vec}(\a Y) } = \Sig_2 \otimes \Sig_1$. In this parameterization 
$\Sig_1$ and $\Sig_2$ represent covariances among the rows and columns of the 
matrices, respectively. Such a covariance model may provide a stable 
and parsimonious alternative to an unrestricted estimate of 
$\Cov{ {\rm vec}(\a Y) } $, the latter being unstable or even unavailable if 
the dimensions of the sample data matrices are large compared to the sample size. 
The family of matrix normal distributions with separable covariance 
matrices is studied in \citet{dawid_1981}, and an iterative algorithm for 
maximum likelihood estimation is given by \citet{dutilleul_1999}. 
Testing various hypotheses regarding the separability of the covariance 
structure, or the form of the component matrices, is considered 
in \citet{lu_zimmerman_2005,roy_khattree_2005,mitchell_genton_gumpertz_2006} 
among others. 
Beyond the matrix-variate case, 
\citet{galecki_1994} considers a separable covariance 
model for three-way arrays, but where the component matrices 
are assumed to have compound symmetry or
an autoregressive structure. 

In this  article we show that the class of separable covariance models
for random arrays of arbitrary dimension  can be generated with a type of multilinear 
transformation known as the Tucker product \citep{tucker_1964,kolda_2006}. 
Just as a zero-mean multivariate normal vector  with a given covariance matrix can be represented as a linear transformation of a vector of independent, standard normal entries,  in Section 2  we show that 
a normal array with separable covariance structure can be represented by 
a multilinear transformation of an array of independent, standard normal 
entries. As a result, the calculations involved in obtaining 
maximum likelihood and Bayesian estimates  are made simple 
via some basic tools of multilinear algebra. 
In Section 3 we  adapt the iterative algorithm of \citet{dutilleul_1999}
to the array normal model, 
and provide  a conjugate prior distribution and 
posterior approximation algorithm for Bayesian inference. 
Section 4
presents an example data analysis of 
trade volume data between pairs of 30 countries in 6 commodity types over 10 years. A discussion of model extensions and directions for further research follows in Section 5.

\section{Separable covariance via array-matrix multiplication}
\subsection{Array notation and basic operations}
An array of order $K$, or  \emph{$K$-array}, 
is a map from the product space of $K$ index sets 
 to the real numbers. 
The different index sets are referred to as the \emph{modes} of the array. 
The \emph{dimension vector} of 
an array gives the number of elements in each index set. For example, 
for a positive integer $m_1$, a vector in $\mathbb R^{m_1}$ is a one-array with dimension $m_1$. 
A matrix in $\mathbb R^{m_1\times m_2}$  is a two-array with 
dimension $(m_1,m_2)$. 
A $K$-array $\a Z$ with dimension $(m_1,\ldots, m_K)$  has elements
$\{ z_{i_1,\ldots, i_K}:  i_k\in\{ 1,\ldots, m_k\},k=1,\ldots, K\}$.

Array \emph{unfolding} refers to the representation of an array by an array of lower order via  combinations of various index sets of an array.
A useful unfolding is the $k$-mode matrix unfolding, or 
$k$-mode \emph{matricization}
\citep{delathauwer_demoor_vandewalle_2000},
in which a $K$-array $\a Z$ 
is reshaped to form a matrix $\a Z_{(k)}$ with $m_k$ 
rows and $\prod_{j:j\neq k} m_k$ 
columns.  Each column corresponds to the entries of $\a Z$ in which the 
$k$th index $i_k$ varies from 1 to $m_k$ and the remaining indices are fixed. 
The assignment of the  remaining indices $\{ i_j: j\neq k\}$ to  
columns of $\a Z_{(k)}$  is determined by the following 
ordering on index sets: 
Letting 
$\v i= (i_1,\ldots, i_K)$ and $\v j =(j_1,\ldots, j_K)$ be two sets  
of indices, 
we say $\v i < \v j$ if $i_k < j_k$ for some $k$ and $i_l \leq j_l$ for
all $l>k$. In terms of ordering the columns of the matricization, this 
means that the index corresponding to a lower-numbered mode ``moves faster''
than that of a higher-numbered mode.

\citet{delathauwer_demoor_vandewalle_2000} define an array-matrix product 
via  the usual matrix product as applied to matricizations. 
The $k$-mode product of an $m_1\times \cdots \times m_K$ 
array $\a Z$ and a $n \times m_k$ matrix $\m A$ is obtained 
by forming the $m_1\times \cdots \times m_{k-1}  \times n \times 
  m_{k+1} \times  \cdots \times m_K$ array 
 from the inversion of the $k$-mode matricization operation
on the matrix 
$\m A\a Z_{(k)} $. The resulting array is denoted  by $\a Z \times_k \m A$. 
Letting  $\m F$ and $\m G$ be matrices of the appropriate sizes, 
important properties of this product include the following:
\begin{itemize}
\item $(\a Z \times_j \m F) \times_k \m G  = 
       (\a Z \times_k \m G) \times_j \m F =  
      \a Z \times_j \m F \times_k \m G$
\item  $(\a Z \times_j \m F)   \times_j \m G = 
   \a Z \times_j (\m G \m F )$
\item $\a Z \times_j ( \m F+ \m G) = \a Z \times_j \m F + \a Z\times_j \m G$. 
\end{itemize}
\citep{delathauwer_demoor_vandewalle_2000}. 
A useful extension of the $k$-mode product is the product of 
an array $\a Z$ with each matrix in a list ${\m A} = \{ \m A_1,\ldots, 
 \m A_K\}$ in which $\m A_k \in \mathbb R^{n_k \times m_k}$, given by 
\[   \a Z \times \a A  =
    \a Z \times_1 \m A_1 \times_2 \cdots \times_K \m A_K . \]
This has been called the ``Tucker operator'' or ``Tucker product'', 
\citep{kolda_2006}, 
 named after the Tucker 
decomposition for multiway arrays \citep{tucker_1964, tucker_1966},
and is used for a type of 
multiway singular value decomposition
 \citep{delathauwer_demoor_vandewalle_2000}. 
A useful calculation involving the Tucker operator is that 
if 
$\a Y = \a Z \times {\m A} $, then  
\[ \a Y_{(k)} = \m A_k \a Z_{(k)} ( \m A_K \otimes \cdots 
\otimes \m A_{k+1} \otimes  \m A_{k-1} \otimes \cdots \otimes \m A_1 ).\] 
Other properties of the Tucker product can be found in 
\citet{delathauwer_demoor_vandewalle_2000} and \citet{kolda_2006}.

\subsection{Separable covariance via the Tucker product}
Recall that the general linear group ${\rm GL}_{m}$ of nonsingular real matrices $\m A$ acts transitively on the 
space ${\rm S}_m$ of positive definite $m\times m$ 
matrices $\Sig$  via the transformation $\m A \Sig  \m A^T$. 
It is convenient to think of 
${\rm S}_m$ as the set of covariance matrices $\{\Cov{\m A \v z}: 
\m A \in {\rm GL}_{m}\} $ where $\v z$ is an $m$-variate 
mean-zero random vector with identity covariance matrix. 
Additionally, if $\v z$ is a vector of independent standard normal 
random variables, then the distributions of $\v y = \m A \m z$ as 
$\m A$ ranges over ${\rm GL}_m$ constitute the family of mean-zero 
vector-valued multivariate normal distributions, which we write as 
$\v y \sim {\rm vnorm}( \v 0 , \Sig)$. 

Analogously, let ${\m  A} = \{ \m A_1,\m A_2\} \in {\rm GL}_{m_1,m_2}
   \equiv {\rm GL}_{m_1} \times {\rm GL}_{m_2}$,  and 
let 
$\m Z$ be a 
$m_1\times m_2$ random matrix with uncorrelated mean-zero variance-one 
entries. The covariance structure of the random matrix $\m Y = \m A_1 \m Z 
 \m A_2^T$ can be described by the $m_1 \times m_1\times m_2 \times m_2$ 
covariance array $\Cov{\a Y }$ 
for which the $(i_1,i_2,j_1,j_2)$ entry is equal to 
$\Cov{ y_{i_1,j_1}, y_{i_2,j_2} }$.  
It is straightforward to show that 
 $\Cov{\m Y }=\Sig_1 \circ \Sig_2$, where $\Sig_j = \m A_j \m A_j^T,\ j=1,2$
and  ``$\circ$'' denotes the outer product. 
This is referred to as a ``separable'' covariance structure, in which the covariance
among elements of $\m Y$ can be described by the row covariance $\Sig_1$ 
and the column covariance $\Sig_2$.  
Letting ``tr()'' be matrix trace and ``$\otimes$'' the Kronecker product, 
well-known alternative ways to describe the covariance structure are as follows:
\begin{eqnarray}
\Exp{ \m Y \m Y^T } &=& \Sig_1 \times {\rm tr}(\Sig_2)  \label{eq:dsepmat} \\
\Exp{ \m Y^T \m Y } &=& \Sig_2 \times {\rm tr}(\Sig_1)  \nonumber \\
\Cov{ {\rm vec}(\m Y) } &=& \Sig_2 \otimes \Sig_1  \nonumber. 
\end{eqnarray}

As $\{ \m A_1 ,\m A_2\}$ ranges over ${\rm GL}_{m_1,m_2}$ the covariance 
array of $\m Y = \m A_1 \m Z\m A_2^T$ ranges over the space of separable 
covariance arrays ${\rm S}_{m_1,m_2} = \{ \Sig_1\circ \Sig_2 : 
   \Sig_1\in {\rm S}_{m_1} ,\Sig_2 \in {\rm S}_{m_2} \}$ 
\citep{browne_shapiro_1991}. 
If we additionally assume that the elements of $\m Z$ are independent 
standard normal random variables, then the distributions of 
$\{ \m Y =  \m A_1 \m Z \m A_2^T :  \{ \m A_1 ,\m A_2 \}\in 
{\rm GL}_{m_1,m_2}\}$ constitute what are known as the mean-zero 
matrix normal distributions \citep{dawid_1981}, which we write as 
$\m Y \sim {\rm mnorm}(\m 0, \Sig_1\circ \Sig_2)$.

Thinking of the matrices $\m Y$ and $\m Z$ as two-way arrays, 
the bilinear transformation $\m Y =  \m A_1 \m Z \m A_2^T$ can alternatively
be expressed using array-matrix  multiplication as 
$\m Y = \m Z \times_1 \m A_1 \times_2 \m A_2 = 
\a Z \times \a A$. 
Extending this idea further, 
let $\a Z$ be an $m_1\times \cdots \times m_K$ random array with uncorrelated mean-zero variance-one entries, 
and define ${\rm GL}_{m_1,\ldots,m_K}$ to be the set of 
lists of matrices ${\m  A} = \{ \m A_1,\ldots, \m A_K \}$ with 
$\m A_k \in {\rm GL}_{m_k}$. The Tucker product 
$\a Z \times  \a A$ induces a transformation 
on the covariance structure of $\a Z$ which shares many features
of the analogous bilinear transformation for matrices:
\begin{prop}  Let $\a Y = \a Z \times {\a  A} $, 
where $\m Z$ and $\tilde {\m A}$ are as above, and let 
$\Sig_k = \m A_k \m A_k^T$.  Then 
\begin{enumerate}
\item $  \Cov{\a Y }  =  \Sig_1 \circ \cdots \circ \Sig_K,  $ 
\item $\Cov{{\rm vec}(\a Y) } = \Sig_K \otimes \cdots  \otimes \Sig_1$, 
\item $\Exp{ \a Y_{(k)}\a Y_{(k)}^T } = \Sig_k \times \prod_{j:j\neq k} {\rm tr}
(\Sig_j)$. 
\end{enumerate}
\end{prop}

The following result highlights the relationship between array-matrix multiplication and separable covariance structure:
\begin{prop}
If  $\Cov{\a Y }=\Sig_1\circ \cdots \circ \Sig_K$ and  $\a X = \a Y \times_k \m G$, then 
 \[  \Cov{\a X }  =  \Sig_1 \circ \cdots \circ \Sig_{k-1} \circ 
(\m G \Sig_k \m G^T )  \circ \Sig_{k+1} \circ \cdots \circ  \Sig_K. \]
\end{prop}
This  indicates that the class of separable covariance 
arrays can be obtained by repeated single-mode array-matrix 
multiplications starting with an array $\a Z$ of uncorrelated entries, i.e.\ for which 
$\Cov{\a Z} = \m I_{m_1} \circ \cdots \circ \m I_{m_K}$. 
The class of 
separable covariance arrays is therefore closed under this group of transformations. 

\section{Covariance estimation with array normal distributions} 
\subsection{Construction of an array normal class of distributions}
Normal probability distributions are useful statistical modeling tools that can represent 
mean and covariance structure. A family of normal distributions 
for random arrays
with separable covariance structure can be generated as in the vector and matrix
cases: Let $\a Z$ be an array of independent standard normal entries,
and let $\a Y = \a M + \a Z \times {\m A} $ with
$\a M \in \mathbb R^{m_1\times \cdots \times m_K}$ and  ${\m A} \in {\rm G
L}_{m_1,\ldots, m_K}$. We say
that $\a Y$ has an array normal distribution, denoted
$\a Y \sim {\rm anorm}(\a M , \Sig_1\circ\cdots\circ \Sig_K)$, where
$\Sig_k = \m A_k \m A_k^T$.

\begin{prop}
The probability density of $\a Y$ is given by
\[  p(\a Y  | \m M, \Sig_1,\ldots, \Sig_K  ) = 
   (2\pi)^{-m/2} 
\left ( \prod_{k=1}^K |\Sig_k|^{-m/(2m_k)}  \right ) \times 
 \exp( - || (\a Y - \a M)\times \Sig^{-1/2} ||^2/2 ), 
 \]
\end{prop}
where   $m = \prod_1^K m_k$, 
$\Sig^{-1/2}  = \{ \m A_1^{-1} ,\ldots, \m A_K^{-1}  \} $ 
and 
the array norm  $||\a Z||^2 = \langle \a Z , \a Z \rangle $ is 
derived from the inner product $\langle \a X, \a Y \rangle  = 
 \sum_{i_1} \cdots \sum_{i_K} x_{i_1,\ldots, i_K } y_{i_1,\ldots, i_K}$.

Also important for statistical modeling is the idea of replication.
If $\a Y_1,\ldots, \a Y_n \iid  {\rm anorm}( \a M, \Sig_1\circ \cdots \circ \Sig_{K})$,
then the array $m_1\times \cdots \times m_K\times n$ array formed by
stacking the $\a Y_i$'s together also has an array normal distribution:
If $\a Y_1,\ldots, \a Y_n \iid  {\rm anorm}( \a M, \Sig_1\circ \cdots\circ \Sig_{K})$, then
\[ \a Y = ( \a Y_1,\ldots, \a Y_{m_K} ) \sim  
  {\rm anorm}(\a M \circ \v 1_n ,\Sig_1\circ \cdots\circ \Sig_{K}\circ \m I_{n}). \]
This can be shown by computing the joint density of
$\a Y_1,\ldots, \a Y_n$ and comparing it to the array normal density.

An important feature of the multivariate normal distribution is that it
provides a conditional model of one set of variables given another. 
Recall, if $\v y \sim$ vnorm$(\v \mu,\Sig)$ then  
the conditional distribution of one subset of elements 
$\v y_b$ of $\v y$ given another  $\v y_a$  is vnorm$( \v\mu_{b|a},
 \Sig_{b|a})$, where 
\begin{eqnarray*}
\v \mu_{b|a} &=& \v \mu_{[b]} +\Sig_{[b,a]} ( \Sig_{[a,a]} )^{-1} (\v y_a - \v \mu_{[a]}) \\
\Sig_{b|a} &=& \Sig_{[b,b]} - \Sig_{[b,a]} ( \Sig_{[a,a]} )^{-1} \Sig_{[a,b]}, 
\end{eqnarray*}
with $\Sig_{[b,a]}$, for example, being the matrix made up of the 
entries in the rows of $\Sig$ corresponding to $b$ and columns 
corresponding to $a$. 

A similar result holds for the array normal distribution: 
Let $a$ and $b$ be non-overlapping subsets of 
$\{1,\ldots, m_1\}$. 
Let $\a Y_{b} = \{ y_{i_1,\ldots, i_K} : i_1 \in \v b\}$
and $\a Y_{a} = \{ y_{i_1,\ldots, i_K} : i_1 \in \v a\}$ 
be arrays of dimension $m_{1b} \times m_2 \times \cdots \times m_K$
and  $m_{1a} \times m_2 \times \cdots \times m_K$, where 
$m_{1a}$ and $m_{1b}$ are the lengths of $a$ and $b$ respectively. 
The arrays $\a Y_{a}$ and $\a Y_b$ are made up of 
non-overlapping  ``slices'' of the array $\a Y$ along the first mode. 
\begin{prop}
  \label{prop:ancond}
Let $\a Y\sim $ {\rm anorm}$( \a M, \Sig_1\circ \cdots\circ \Sig_K)$. 
The conditional distribution of $\a Y_b$ given $\a  Y_a$ is 
array normal  with mean $\a M_{b|a}$ and covariance  
$\Sig_{1,b|a}\circ\Sig_2\circ \cdots\circ\Sig_K$, where 
\begin{eqnarray*}
 \a  M_{b|a} &=& \a M_b +  (\a Y_{a}-\a M_a ) \times_1 ( \Sig_{1[b,a]} (\Sig_{1[a,a]})^{-1} )  \\ 
\Sig_{b|a} &=& \Sig_{1[b,b]} - \Sig_{1[b,a]} ( \Sig_{1[a,a]} )^{-1} \Sig_{1[a,b]}.  
\end{eqnarray*} 
\end{prop}
Since the conditional distribution is also in the array normal class, 
successive applications 
of Proposition \ref{prop:ancond} can be used to obtain 
the conditional distribution of  any subset of the elements of 
$\a Y$  of the form $\{ y_{i_1,\ldots, i_K} : i_k \in b_k \}$, 
conditional upon the other elements of the array. 


\subsection{Estimation for the array normal model}
\paragraph{Maximum likelihood estimation:}
Let $\a Y_1,\ldots, \a Y_n \iid {\rm anorm}(\a M, \Sig_1\circ \cdots \circ \Sig_K)$, or equivalently, 
 $\a Y \sim {\rm anorm}( \a M\circ \v 1 , \Sig_1\circ \cdots \circ \Sig_K \circ \m I_n )$.   For any value of $\Sig = \{\Sig_1,\ldots, \Sig_K\}$, the 
value of $\a M$ that maximizes $p(\a Y | \a M , \Sig)$ is the value that 
minimizes the residual mean squared error:
\begin{eqnarray*}
 \frac{1}{n}\sum_{i=1}^n|| (\a Y_i-\a M ) \times \Sig^{-1/2} ||^2 &=&  
\frac{1}{n}  \sum_{i=1}^n \langle \a Y_i -\a M , (\a Y_i -\a M)\times \Sig^{-1} \rangle \\
&=&  \langle \a M , \a M \times \Sig^{-1} \rangle   - 
    2 \langle \a M , \bar {\a  Y } \times \Sig^{-1} \rangle  + c_1(\a Y,\Sig)  
  \\
&=&  \langle \a M- \bar{\a Y } , (\a M -\bar{\a Y} ) \times \Sig^{-1} \rangle
  +c_2(\a Y,\Sig) \\
&=& || (\a M -\bar{\a Y} ) \times \Sig^{-1/2} ||^2 + c_2(\a Y,\Sig). 
\end{eqnarray*}
This is uniquely minimized in $\m M$ by $\bar {\a Y } = \sum \a Y_i/n$, 
and so $\hat {\m M} = \bar{\a Y} $ is the MLE of $\a M$. 
The MLE of $\Sig$ does not have a closed form expression. However, it is 
possible to maximize  $p(\a Y | \a M , \Sig)$  in $\Sig_k$, given 
values of the other covariance matrices.
Letting $\a E = \a Y -  \a M\circ \v 1_n$, 
the likelihood  as a function of $\Sig_k$ can be expressed as 
$p(\a Y | \a M, \Sig )  \propto  
 |\Sig_k|^{-nm/(2m_k)}  \exp\{ -|| \a E\times \{ \Sig_1^{-1/2},
  \ldots, \Sig_K^{-1/2},\m I_n\} ||^2  /2 \} $. 
Since for any array $\m Z$ and mode $k$ we have  $||\a Z ||^2 = || \m Z_{(k)}||^2 = \tr( \a Z_{(k)}\a Z_{(k)}^T)$, 
the norm in the likelihood can be written as 
\begin{eqnarray*}
|| \a E  \times \Sig^{-1/2} ||^2 &=& || \tilde {\a E}   \times_k \Sig_k^{-1/2}||^2  \\
&=& \tr ( \Sig_k^{-1/2} \tilde{ \a E}_{(k)} \tilde{ \a E}_{(k)}^T \Sig_k^{-1/2} )  \\
&=& \tr ( \Sig_k^{-1} \tilde{ \a E}_{(k)} \tilde{ \a E}_{(k)}^T ),
\end{eqnarray*}
where $\tilde{\a E} =\a E \times \{ \Sig_1^{-1/2},\ldots, 
\Sig_{k-1}^{-1/2},\m I_k  , \Sig_{k+1}^{-1/2},\ldots, \Sig_K^{-1/2},\m I_n\}$
is the residual array standardized along each dimension except $k$. 
Writing $\m S_k = \tilde{ \a E}_{(k)} \tilde{ \a E}_{(k)}^T $, we have 
\begin{eqnarray*}
p(\m Y | \m M,  \m \Sig ) & \propto & 
|\Sig_k|^{-nm/(2m_k)} \etr\{-\Sig_k^{-1} \m S_k/2 \}
\end{eqnarray*}
as a function of $\Sig_k$, 
and so if $\m S_k$ is of full rank then 
the unique maximizer in $\Sig_k$ is given by 
$\hat {\Sig}_k  =  \m S_k/n_k$, where 
$n_k = nm/m_k =  n \times \prod_{j\neq k} m_j$ is the  number of columns of 
$\a Y_{(k)}$, i.e.\  
the ``sample size'' for the 
$k$th mode. 
This suggests the following  iterative algorithm for obtaining 
the MLE of $\Sig$: Letting $\a E = \a Y - \bar {\a Y} \circ \v 1_n $ and 
given an initial value of  $\Sig$,  for each $k\in \{1,\ldots, K\}$
\begin{enumerate}
\item compute $\tilde{\a E} =\a E \times \{ \Sig_1^{-1/2},\ldots, 
\Sig_{k-1}^{-1/2},\m I_k  , \Sig_{k+1}^{-1/2},\ldots, \Sig_K^{-1/2},\m I_n\}$ 
and $\m S_k = \tilde{ \a E}_{(k)} \tilde{ \a E}_{(k)}^T $; 
\item set $\Sig_k  = \m S_k/n_k$, where $n_k = n\times \prod_{j\neq k} m_j$. 
\end{enumerate}
Each iteration increases the likelihood, and so the procedure can be seen 
as a type of block coordinate descent algorithm \citep{tseng_2001}. 
For the matrix normal case, this algorithm was proposed by 
\citet{dutilleul_1999} and is sometimes called the ``flip-flop'' algorithm. 
Note that  the scales of $\{\Sig_1,\ldots, \Sig_K\}$ are
not separately identifiable from the likelihood: Replacing
$\Sig_1$ and $\Sig_2$ with $c\Sig_1 $ and $\Sig_2/c$ yield the
same probability distribution for $\a Y$, and so the scales of  
the MLEs will depend on the initial values.

\paragraph{Bayesian estimation:}
Estimation of high-dimensional parameters
often benefits from a complexity penalty, e.g.\ a penalty on the 
magnitude of the parameters. Such penalties can often be expressed as 
prior distributions, and so penalized likelihood estimation can be done 
in the context of Bayesian inference. With this in mind, we consider 
semiconjugate prior distributions for the array normal model, and their 
associated posterior distributions. 

A conjugate prior distribution for the 
for the multivariate normal model $\v y_1\ldots \v y_n \iid $ 
vnorm$(\v \mu,\Sig)$ 
is given by $p(\v \mu ,\Sig ) =  p(\v \mu|\Sig)  p(\Sig)$, where  
$p(\Sig)$ is an inverse-Wishart density and 
$ p(\v \mu|\Sig)$ is multivariate normal density with 
prior mean $\v \mu_0$ and prior (conditional) covariance 
$\Sig/\kappa_0$. The parameter $\kappa_0$ can be thought of as a 
``prior sample size,'' as the the prior covariance $\Sig/\kappa_0$ 
for $\v\mu$ is the same as that of a sample average based on $\kappa_0$ 
observations. Under this prior distribution, the 
conditional distribution of $\v \mu $ given the data and $\Sig$ is  
multivariate normal, and the condition distribution of $\Sig$  given the data 
is 
inverse-Wishart. 
An analogous result holds for the array normal model:
If 
\begin{eqnarray*}
\a M | \Sig &\sim &  \mbox{anorm}( \a M_0 , \Sig_1\circ \cdots \circ \Sig_K/\kappa_0) \\
\Sig_k &\sim&  \mbox{inverse-Wishart} (  \m S_{0k}^{-1} ,\nu_{0k}) 
\end{eqnarray*}
and $\Sig_1,\ldots, \Sig_K$ are independent, then  straightforward 
calculations show that 
\begin{eqnarray*}
\a M | \a Y,  \Sig &\sim &  \mbox{anorm}( [\kappa_0 \a M_0+n \bar {\a Y}]/[\kappa_0+n]  , \Sig_1\circ \cdots \circ \Sig_K/[\kappa_0+n ]) \\
\Sig_k | \a Y, \Sig_{-k}& \sim &   \mbox{inverse-Wishart} (  [\m S_{0k}+\m S_k  + \m R_{(k)} \m R_{(k)}^T ]^{-1}  ,\nu_{0k} + n_k ) , 
\end{eqnarray*}
where $\m S_k$ and $n_k$ are as in the coordinate descent algorithm for 
maximum likelihood estimation, and 
$\m R =\sqrt{\frac{\kappa_0 n}{\kappa_0+n}} (\bar {\a Y}-\a M_0) \times \{  
 \Sig_1^{-1/2} ,\ldots, \Sig_{k-1}^{-1/2} ,\m I , \Sig_{k+1}^{-1/2} ,\ldots, 
\Sig_K^{-1/2} \} $.

As noted above, the scales of $\{\Sig_1,\ldots, \Sig_K\}$ are 
not separately identifiable from the likelihood. 
This makes the 
prior and posterior distributions of the scales of the $\Sig_k$'s difficult 
to specify or interpret. As a remedy, we consider reparameterizing 
the  prior
distribution for $\Sig_1,\ldots, \Sig_K$ to include a parameter 
representing the total variance in the data. Parameterizing 
 $\m S_{0k} = \gamma \Sig_{0k}$ for each $k\in \{1,\ldots, K\}$, 
the prior expected 
total variation of $\a Y_i$, ${\rm tr}( \Cov { {\rm vec}( {\a Y_i})})$, 
is 
\begin{eqnarray*} 
\Exp{ {\rm tr}( \Cov { {\rm vec}( {\a Y})}) }  &=& 
\Exp{ {\rm tr}( \Sig_K \otimes \cdots \otimes  \Sig_1 ) } \\
 &=& \Exp{ \prod_{k=1}^K {\rm tr}( \Sig_k) }  \\ 
 &=& \prod_{k=1}^K {\rm tr}( \Exp{\Sig_k})  
=   \gamma^K \prod_{k=1}^K {\rm tr}(\Sig_{0k})/(\nu_{0k}-m_k-1 ) . 
\end{eqnarray*}
A simple default choice for $\Sig_{0k}$ and $\nu_{0k}$ would be 
 $\Sig_{0k}= \m I_{m_k}/m_k$ and $\nu_{0,k} = m_k +2$, for which 
$\Exp{ {\rm tr}(\Sig_k) } = \gamma$ and the expected value for the 
total variation is $\gamma^K$. Given prior expectations about the total 
variance, the value of $\gamma$ could be set accordingly. Alternatively, 
a prior distribution could be placed on $\gamma$: If 
$\gamma \sim {\rm gamma}(a,b)$ with prior mean $a/b$, 
then conditional on $\Sig_1,\ldots, \Sig_K$, we have 
\[ \gamma|\Sig_1,\ldots \Sig_K \sim 
 {\rm gamma}\left ( a+ \sum \nu_{0k} m_k /2 , b+ \sum {\rm tr}(\Sig_k^{-1} \Sig_{0k} )/2 \right  ). \]

The full conditional distributions of $\{\a M , \Sig_1,\ldots, \Sig_K, \gamma\}$ can be used to implement a Gibbs sampler, in which each parameter is 
 sampled  in turn from its full conditional distribution, given the 
current values of the other parameters.  This algorithm generates 
a Markov chain having a stationary density equal to 
$p(\a M , \Sig_1,\ldots, \Sig_K, \gamma | \a Y )$, samples from which 
can be used to approximate posterior quantities of interest. 
Such an algorithm is implemented in the data analysis example in the next section. 

\section{Example: International trade}
The United Nations gathers yearly trade data between 
countries  of the world and disseminates this information at the 
UN Comtrade website \href{http://comtrade.un.org/}{\nolinkurl{http://comtrade.un.org/}}.  In this section we analyze trade among pairs of countries over several years and in several different commodity categories. 
Specifically, the data take the form of a four-mode array $\m Y = \{ 
y_{i,j,k,t} \}$ where 
\begin{itemize}
\item $i \in \{1,\ldots, 30=m \}$ indexes the exporting nation; 
\item $j \in \{1,\ldots, 30=m \}$ indexes the exporting nation; 
\item $k \in \{1,\ldots, 6 =p \}$ indexes the commodity type; 
\item $t \in \{1,\ldots, 10 = n\}$ indexes the year. 
\end{itemize}
The thirty countries were selected to make the data as complete as possible, 
resulting in a set of 
mostly large or developed countries with high gross domestic 
products and trade volume.  The six commodity types include
(1) chemicals,
(2) inedible crude materials not including fuel,
(3) food and live animals, 
(4) machinery and transport equipment, 
(5) textiles and (6) manufactured goods. 
The years represented in the dataset include 1996 through 2005. 
As trade between countries is relatively stable across years, 
we analyze the yearly change in log trade values, measured in 2000 US dollars. 
For example, $y_{1,2,1,1}$ is the log-dollar increase in the 
value of chemicals  exported 
from Australia to Austria  from 1995 to 1996. 
We note that exports of a country to itself are not defined, so 
$y_{i,i,j,t}$ is ``not available''  and can be treated as missing 
at random. 

We model these data as $\a Y = \a M \circ \v 1_n  + \a E $ , 
where $\a M$ is an $m\times m \times p$ array of means specific to 
 exporter-importer-commodity combinations, and $\a E$ is an 
$m\times m\times p\times n$ array of residuals.  Of interest here 
is how the deviations  $\a E$ of the data from the mean may be correlated 
across exporters, importers and commodities. 
One possible model for this residual variation
would be to treat the  $p$-dimensional residual vectors 
corresponding to each of the $m\times (m-1) \times n=8700$ 
exporter-importer-year
combinations as independent samples from a $p$-variate multivariate normal 
distribution. 
However, to accommodate potential temporal correlation 
(beyond that already accounted for by taking $\a Y$ to be the lagged log trade values), 
the $p\times n$ residual matrices corresponding to each of the 
$m\times (m-1) = 870$ exporter-importer pairs could
be modeled as independent 
samples from a 
matrix normal distribution, with two separate covariance matrices representing 
commodity and temporal correlation.  This latter model can be described 
by an array normal model as 
\begin{equation} 
\a Y \sim \mbox{anorm}( \a M \circ \v 1_n , \m I_m \circ \m I_m \circ \m \Sig_3 \circ \Sig_4 ),  
\label{eq:trade.red}
\end{equation}
where $\Sig_3$ and $\Sig_4$ describe covariance among commodities and time points, respectively. However, it is natural to consider the possibility that 
there will be 
correlation of residuals attributable to exporters and importers. For 
example, countries with similar economies may  exhibit 
correlations in their trade patterns.  With this in mind, we will also
fit the following model:
\begin{equation} \a Y \sim \mbox{anorm}( \a M \circ \v 1_n , \m \Sig_1 \circ \m \Sig_2 \circ \m \Sig_3 \circ \Sig_4 ). 
\label{eq:trade.full}
\end{equation}
We consider Bayesian analysis for both of these models based on the prior 
distributions described at the end of the last section. 
The prior distribution for each $\Sig_k$ 
matrix being estimated is given by  $\Sig_k  \sim$ inverse-Wishart$( m_k \v I_{m_k}/\gamma, m_k+2)$, with 
the hyperparameter $\gamma$ set so that $\gamma^K = || \a Y - \bar {\a Y}\circ\v 1_n ||^{2}$.  As described in the previous section, this weakly centers the 
total variation of $\a Y$ under the model around the the empirically observed value, similar to an empirical Bayes approach or unit information prior distribution \citep{kass_wasserman_1995}. 
The prior distribution for $\m M$ conditional on 
$\Sig$  is 
$\a M \sim$ anorm$(\a 0 , \Sig_1\circ \Sig_2 \circ \Sig_3)$, 
where $\Sig_1=\Sig_2= \m I_m $ for model \ref{eq:trade.red}. 

Posterior distributions of parameters  for both model \ref{eq:trade.red} and
\ref{eq:trade.full} can be obtained using the results of the previous section
with minor modifications.
Under both models
the $m\times m\times p$ arrays $\a Y_1,\ldots, \a Y_{n}$ 
corresponding to the $n=10$ time-points are not independent, but correlated 
according to  $\Sig_4$. 
The full conditional distribution of $\a M$ is still given by an 
array normal distribution, but the mean and variance are now as 
follows:
\begin{eqnarray*}
\Exp{ \a M | \a Y , \Sig } &=& 
\frac{ \kappa_0 \a M_0 + \sum_{i=1}^n  c_i  \tilde {\a Y}_i }{\kappa_0 +  
 \sum c_i^2 }  \\
\Var{ \a M | \a Y ,  \Sig } &=& 
\Sig_1\circ \Sig_2 \circ \Sig_3 \circ \Sig_4/(\kappa_0 + \sum c_i^2  )
\end{eqnarray*}
where $\tilde{\a Y}_1,\ldots, \tilde {\a Y}_n$ are the $m\times m \times p$ 
arrays obtained from the first three modes of the transformed array 
$\tilde {\a Y }  =  \a Y \times_4 \Sig^{-1/2}_4 $, and 
$c_1,\ldots, c_n$ are the elements of the vector
$\v c = \Sig_4^{-1/2} \v 1 $. 
Additionally,  the time dependence makes it difficult to integrate 
$p( \a M , \Sig|\a Y)$ as was possible in the independent case. 
As a result,  we use a 
Gibbs sampler that proceeds by sampling 
$\Sig_k$ from its full conditional distribution $p(\Sig_k| \a Y , \a M, \Sig_{-k})$ as opposed to the $p(\Sig_k| \a Y , \Sig_{-k})$ as before. This full conditional distribution is still a member of the inverse-Wishart family:
\begin{eqnarray*}
\Sig_k | \a Y,\a M, \Sig_{-k}& \sim &   \mbox{inverse-Wishart} (  [\m S_{0k}+\m E_{(k)} \m E_{(k)}^T  + \m R_{(k)} \m R_{(k)}^T ]^{-1}  ,\nu_{0k} + n_k\times [1+1/n] ) , 
\end{eqnarray*}
where $\a E_{(1)}$, for example, is the $k$-mode matricization of 
$(\a Y-\a M\circ \v 1_n) \times \{  
 \m I_m ,\Sig_{2}^{-1/2} ,\Sig_{3}^{-1/2} ,\Sig_4^{-1/2} \}$
and $\a R_{(1)}$ is the $k$-mode matricization of 
$ (\a M-\a M_0) \times \{  \m I ,\Sig_{2}^{-1/2} ,\Sig_{3}^{-1/2} \}$. 

Separate Markov chains for each of the two models 
were generated using 205,000 iterations of the Gibbs sampler discussed above. 
The first  5,000  iterations were dropped from each chain to allow 
for convergence to the stationary distribution, and parameter values were 
saved every 40th iteration thereafter, resulting in  5,000 parameter values 
with which to approximate the posterior distributions. 
Mixing of the Markov chain was assessed by computing the ``effective sample 
size'', or equivalent number of independent simulations, of several summary 
parameters. For the full model,  effective sample sizes of 
$\gamma_0=\tr( \Sig_1\otimes \cdots \otimes \Sig_4), \gamma_1=\tr(\Sig_1),\ldots, \gamma_4 =\tr(\Sig_4)$   were computed to be 2,545, 904, 960, 548 and 1,734.
Note that $\gamma_1,\ldots, \gamma_4$ are not separately identifiable 
from the data, resulting in poorer mixing than $\gamma_0$, which is 
identifiable. For the reduced model, the effective sample sizes of 
$\gamma_0$, $\gamma_3$ and $\gamma_4$ were 4,281, 1,194 and 1,136.

\begin{figure}[ht]
\centerline{\includegraphics[width=6.5in]{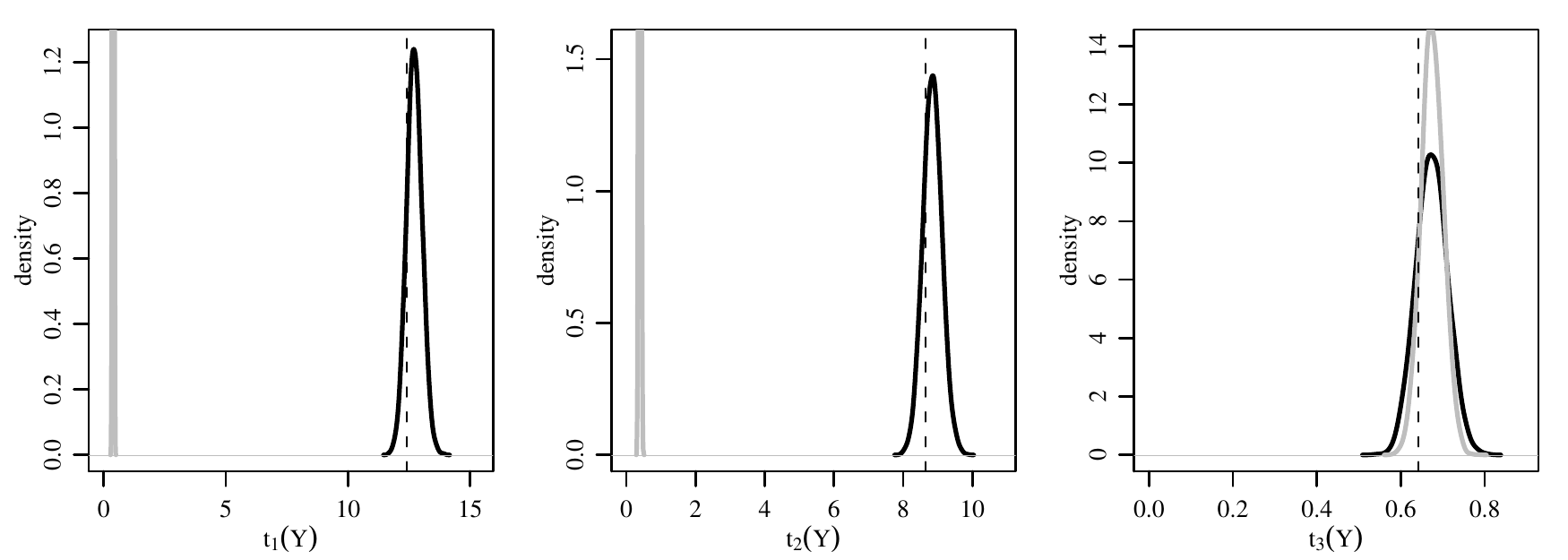}}
\caption{Posterior predictive distributions for summary statistics. The gray density represents is under the reduced model, the black under the full. The vertical dashed line is the observed value of the statistic.  }
\label{fig:fig1}
\end{figure}

\begin{figure}[ht]
\centerline{\includegraphics[width=6.75in]{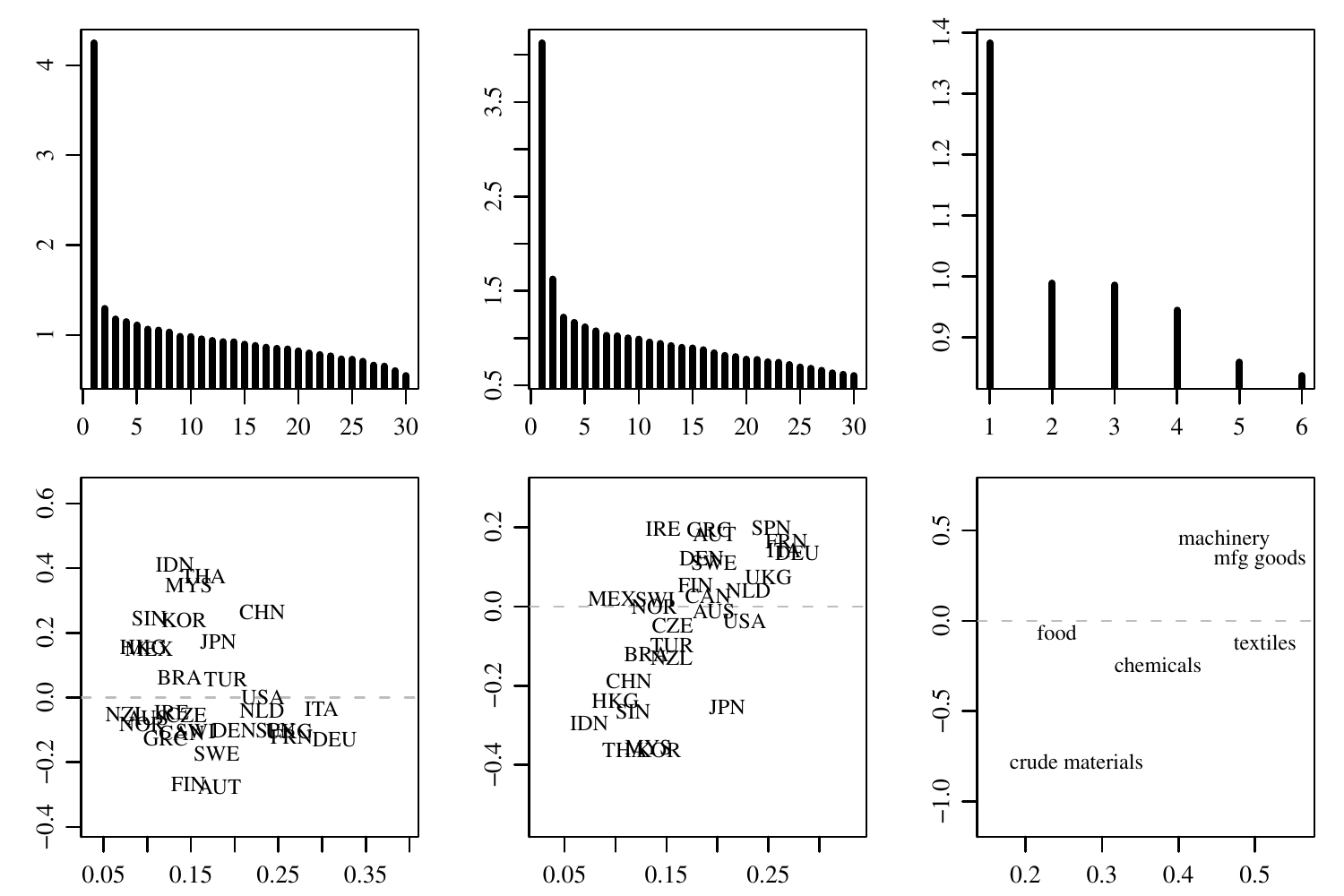}} 
\caption{Estimates of correlation matrices corresponding to
$\Sig_1$, $\Sig_2$ and $\Sig_3$. The first panel in each row plots
the eigenvalues of each correlation matrix, the second plots the first two eigenvectors. }
\label{fig:fig2}
\end{figure}

The fits of the two models can be compared using posterior predictive 
evaluations \citep{rubin_1984}: 
To evaluate the fit of a model,  the observed value 
of a summary statistic $t(\a Y)$ can be compared to 
values $t(\tilde {\a Y })$ for 
which $\tilde {\a Y }$ is simulated from the posterior predictive distribution.
A discrepancy between $t(\a Y)$ and the distribution  
of $t(\tilde {\a Y })$  indicates that the model is not capturing the aspect of 
the data represented by $t(\cdot)$.  For illustration, we use such checks 
here to evaluate evidence that $\Sig_1$ and $\Sig_2$ are not equal to 
the identity, or equivalently, that model 1 exhibits lack of fit as compared 
to model 2. To obtain a summary statistic evaluating evidence of a non-identity covariance 
matrix for $\Sig_1$, we first subtract
the sample mean from the data array to obtain $\a E = \a Y - \bar {\a Y}$, 
and then compute $\m S_1 = (\a E_{(1)} \a E_{(1)}^T)$. The  $m\times m$ matrix $\m S_1$ is a  sample measure  of covariance among exporting countries. 
We then obtain a scaled version $\tilde {\m S}_1 =  \m S_1/\tr(\m S_1)$, 
and compare it to a scaled version of the identity matrix:
\[ t_1(\a Y) = \log |\tilde {\m S}_1| - \log | \m I/m|   = 
   \log |\tilde {\m S}_1  | + m\log m. \]
Note that the minimum value of this statistic occurs when $\tilde {\m S}_1= 
 \m I/m$, and so in some sense it provides a simple scalar measure of how the sample covariance 
among exporters differs from a scaled identity matrix. 
Similarly, we construct $t_2(\a Y)$ and $t_3(\a Y)$ measuring sample 
covariance along the 
second and third modes of the data array. We include 
$t_3$ to contrast with  $t_1$ and $t_2$, as both the full and reduced models 
include covariance parameters for the third dimension of the array.

Figure \ref{fig:fig1} plots posterior predictive densities for 
$t_1(\tilde {\a Y})$, $t_2(\tilde {\a Y}) $ and 
 $t_3(\tilde {\a Y}) $ under both the full and reduced models, and compares 
these densities to the observed values of the statistics. 
The reduced model exhibits substantial lack of fit in terms of its inability 
to predict datasets that resemble the observed in terms of $t_1$ and $t_2$. 
In other words, a model that assumes i.i.d.\ structure along the 
first two modes of the array does not fit the data. In terms of covariance 
among commodities along the third mode, neither model exhibits substantial lack of fit as measured
by $t_3$.

Figure  \ref{fig:fig2} describes posterior mean estimates of the 
correlation matrices corresponding to $\Sig_1$, $\Sig_2$ and $\Sig_3$. 
The two panels in each column plot the eigenvalues and the first 
two eigenvectors each of the three correlation matrices. 
The eigenvalues for all three suggest the possibility of modeling 
the covariances matrices with factor analytic structure, i.e.\ 
letting $\Sig_k =  \m A \m A^T + {\rm diag}(b_1^2,\ldots, b_{m_k}^2)$, 
where $\m A$ is an $m_k\times r$ matrix with $r<m_k$. This possibility is 
described further in the Discussion.  The second row of  Figure 
\ref{fig:fig2}  describes correlations among exporters, importers and 
commodities. The first two plots show that much of the correlation 
among exporters and among importers is 
related to geography, with countries having similar eigenvector values 
typically being near each other geographically. The third plot in the 
row indicates correlation among commodities of a similar type: 
Moving up and to the right from ``crude materials,''  the commodities 
are essentially in order of the extent to which they are finished goods.

\section{Discussion}
This article has proposed a class of array normal distributions with separable 
covariance structure for the modeling of array data, particularly 
multivariate relational arrays. 
It has been shown that the class of separable covariance models 
can be generated by 
a type of array-matrix multiplication 
known as the Tucker product. 
Maximum likelihood estimation and Bayesian inference in Gaussian versions of 
such models are feasible using relatively simple properties of 
array-matrix multiplication. These types of models can be useful 
for describing covariance within the index sets of an array dataset. 
In an example involving longitudinal trade data, 
we have shown that a  full array normal model provides a better 
fit than a
matrix normal model, 
which includes a covariance  matrix
 for only two of the four modes of the data array. 

One open area of research is  finding conditions for 
the existence and uniqueness of the MLE for the array normal model. 
In fact, such conditions for even the simple matrix normal case are 
not fully established.  Each step of the iterative MLE algorithm of 
\citet{dutilleul_1999} is well defined as long as $n \geq \max\{m_1/m_2,m_2/m_1\} +1 $, suggesting that this condition might be sufficient to provide existence.  Unfortunately, it is easy to find examples where this condition is met but the likelihood is 
unbounded, and no MLE exists. In contrast, \citet{srivastiva_vonrosen_vonrosen_2008}   show the MLE exists and is essentially unique if $n> \max\{m_1,m_2\}$, 
a much stronger requirement. However, they do not show that this condition is 
necessary. Unpublished results of this author suggest that this latter
condition can be relaxed, and that the existence of the MLE depends on 
the minimal possible rank of $\sum_{i=1}^n \m Y_i \m U_k\m U_k^T \m Y_i^T$
for $k\leq \min\{m_1,m_2\}$, where $\m U_k$ is an $m_2\times k$ matrix with orthonormal columns. However, such conditions are somewhat opaque, and it is not clear that they could easily be generalized to the array normal case. 

A potentially useful model variation would  be to consider 
imposing simplifying structure on the component matrices. 
For example, a normal
factor analysis model for a random vector $\v y \in \mathbb R^p$ 
posits that
$\v y = \v \mu + \m B \v z +  \m D \v \epsilon$ 
where $\v z\in \mathbb R^r, r<p$ and 
$\v \epsilon  \in \mathbb R^p$ are uncorrelated standard normal vectors
and $\m D$ is a diagonal matrix. 
The resulting covariance matrix is given by
$\Cov{\v y} =  \m B \m B^T + \m D^2$, in which the ``interesting'' part
 $\m B\m B^T$  is of rank $r$.
The natural extension to random arrays is 
$\a Y_i  = \a M + \m Z\times\{ \m B_1,\ldots, \m B_K\} 
+  \m E \times \{ \m D_1,\ldots, \m D_K \} $ where 
$\a Z\in \mathbb R^{r_1 \times \cdots \times r_K}$ 
and $\a E \in \mathbb R^{p_1\times \cdots \times p_K}$ 
are uncorrelated standard normal arrays. 
This induces the covariance matrix 
$\Cov{{\rm vec}(\m Y) } =  (\m B_K \m B_K^T) \otimes \cdots \otimes 
   (\m B_1 \m B_1)^T +    \m D_K^2 \otimes \cdots \otimes \m D_1^2$. 
  This is essentially the model-based analogue of 
the higher-order SVD of \citet{delathauwer_demoor_vandewalle_2000}, in 
the same way that the usual factor analysis model for vector-valued data 
is analogous to the matrix SVD. 
Alternatively, in some cases it may be desirable to 
fit a factor-analytic structure for the covariances of  some modes of the 
array
while estimating others as unstructured. 
This can be achieved with a model of the following form:
\begin{eqnarray*}
\a Y &=& \a M +  \a Z \times \{ \Sig_1^{1/2} ,\ldots, \Sig_k^{1/2},
\m B_{k+1},
\ldots, \m B_{K} \} + \a E \times \{ \Sig_1^{1/2} ,\ldots, \Sig_k^{1/2}, 
  \m D_{k+1}  , \ldots,  \m D_{K} \}
\end{eqnarray*}
where $\a Z\in \mathbb R^{p_1\times \cdots p_k \times r_{k+1} \times  \cdots r_K}$,
and  $\a E\in \mathbb R^{p_1\times \cdots \times p_K}$.
The resulting covariance for  $\a Y$ is given by 
\[ \Cov{{\rm vec}(\m Y) } =  (\m B_K \m B_K^T+ \m D_K^2) \otimes \cdots \otimes 
   (\m B_{k+1} \m B_{k+1}+\m D_{k+1}^2 )^T  \otimes \Sig_k \otimes \cdots \otimes \Sig_1,  \] 
which is separable, and so is in the array normal class. Such a factor model 
may be useful if some modes of the array have very high dimensions,  
and  rank-reduced estimates of  the corresponding 
covariance matrices are desired. 

An additional model extension would be to 
accommodate non-normal continuous or  discrete array data, 
for example, dynamic binary networks.  This can be done by embedding the 
array normal model within a generalized linear model, or within an ordered 
probit model for ordinal response data.  For example, if 
$\a Y$ is a three-way binary array, an array normal probit model would posit
a latent array $\a Z \sim {\rm anorm}( \a M , \Sig_1\circ \Sig_2 \circ \Sig_3)$
which determines $\a Y$ via $y_{i,j,k } = \delta_{(0,\infty) } (z_{i,j,k})$.

Computer code and data for the example  in Section 5 
is available at my website:  \href{http://www.stat.washington.edu/~hoff}{\nolinkurl{http://www.stat.washington.edu/~hoff}}.


\section*{Appendix}
{\sc Proof of Proposition 3.1.}
Let $\a Y =   \a Z \times {\m A}   $ where the 
elements of $\a Z$ are uncorrelated, have expectation zero and 
variance one. 
Using the fact that  
$\m Y_{(1)} = \m A_1 \a Z_{(1)} \m B^T$ where $\m B = ( \a A_K \otimes \cdots \otimes \a A_2)$
 \citep[Proposition 4.3]{kolda_2006}, we have 
\begin{eqnarray*} 
 {\rm vec}(\a Y)  = {\rm vec}(\a Y_{(1)} )  &=& 
     {\rm vec} ( \m A_1 \a Z_{(1)} \a B^T ) \\ 
 &=&  ( \m B  \otimes \m A_1 ) {\rm vec}(\m Z_{(1)}  ) =
   ( \m B  \otimes \m A_1 ) {\rm vec}(\m Z )  .  
\end{eqnarray*}
The covariance of ${\rm vec}(\a Y)$ is then 
\begin{eqnarray*}
\Exp{ {\rm vec}(\a Y){\rm vec}(\a Y)^T } &=& 
(\m B\otimes \m A_1) \Exp{ {\rm vec}(\a Z){\rm vec}(\a Z)^T } (\m B\otimes \m A_1)^T  \\ 
&=& (\m A_K \otimes \cdots \otimes \m A_1) \m I (\m A_K \otimes \cdots \otimes \m A_1)^T \\
 &=&  (\m A_K \m A_K^T) \otimes \cdots \otimes (\m A_1 \m A_1^T)  \\  
 &=& \Sig_K \otimes \cdots \otimes \Sig_1,
\end{eqnarray*}
where $\Sig_k = \m A_k \m A_k^T$.  This proves the second statement in the proposition. The first statement follows from how the ``vec'' operation is applied to arrays. For the third statement, consider the calculation of 
$\Exp{ \m Y_{(1)} \m Y_{(1)}^T} $,  again  using the fact that
 $\m Y_{(1)} = \m A_1 \a Z_{(1)} \m B^T$:
\begin{eqnarray}
\Exp{ \m Y_{(1)} \m Y_{(1)}^T} &=& 
 \m A_1  \Exp{ \m Z_{(1)} \m B^T\m B \m Z_{(1)}^T }  \m A_1^T   \nonumber \\
 &=& \m A_1 \Exp{\m X \m X^T } \m A_1^T,  \label{eq:a1}
\end{eqnarray}
where $\m X = \m Z_{(1)} \m B^T$. Because the elements of $\m Z$ are all 
independent, mean zero and variance one, the rows of $\m X$ are independent 
with mean zero and variance $\m B \m B^T$. 
Thus $\Exp{\m X\m X^T } =  \tr (\m B\m B^T)  \m I$. Combining this with 
(\ref{eq:a1})
gives 
\begin{eqnarray*}
\Exp{ \m Y_{(1)} \m Y_{(1)}^T} 
&=& \m A_1 \m A_1^T  \tr(\m B\m B^T) \\
&=& \m A_1 \m A_1^T \tr(  [\m A_K\otimes \cdots\otimes  \m A_2 ] [ \m A_K^T\otimes \cdots \otimes \m A_2^T ] \\ 
&=& \Sig_1 \tr( \Sig_K \otimes \cdots \otimes \Sig_1 ) \\
&=& \Sig_1 \prod_{k=2}^K \tr( \Sig_k). 
\end{eqnarray*}
Calculation of $\Exp{ \m Y_{(k)} \m Y_{(k)}^T}$ for other values of $k$ 
is similar.   $\Box$

{\sc Proof of Proposition 3.2.}  
We calculate $\Exp{ {\rm vec}(\a X) {\rm vec}(\a X)^T } $ for the 
case that $\a X = \a Y \times_1 \m G$:
\begin{eqnarray*}
 {\rm vec}(\a X) =  {\rm vec}(\a X_{(1)} ) &=& 
   {\rm vec}( \m G\a Y_{(1)} )  \\
 &=&  {\rm vec}( \m G\a Y_{(1)} \m I)  \\ 
 &=& ( \m I \otimes \m G ) {\rm vec}(\a Y) \  ,  \ \mbox{so} \\
\Exp{ {\rm vec}(\a X) {\rm vec}(\a X)^T } &=& 
   ( \m I \otimes \m G ) \Exp{ {\rm vec}(\a Y) {\rm vec}(\a Y)^T } 
    ( \m I \otimes \m G )^T \\
 &=& ( \m I \otimes \m G )  ( \Sig_K \otimes \cdots \otimes \Sig_1) 
     ( \m I \otimes \m G^T ) \\
&=& [  ( \Sig_K \otimes \cdots \otimes \Sig_2) \otimes  (\m G \Sig_1) ] 
    [ \m I \otimes \m G^T ] \\
&=& (\Sig_K\otimes \cdots \otimes \Sig_2 )\otimes  (\m G \Sig_1 \m G^T). 
\end{eqnarray*}
Calculation for the covariance of $\a X = \a Y\times_k  \m G$ for other values of $k$ proceeds analogously. 
$\Box$

{\sc Proof of Proposition 4.1}  
The density can be obtained as a re-expression  of the 
density of $\v e = {\rm vec}(\a E) = {\rm vec}(\a Y-\a M)$, which has a multivariate normal 
distribution with mean zero and covariance $\Sig_K \otimes \cdots \otimes \Sig_1$.  The re-expression is obtained using the following identities, 
\begin{eqnarray*} 
|| (\a Y -\a M )\times \Sig^{-1/2} ||^2 &=& 
  \langle \a E , \a E\times \Sig^{-1} \rangle  \\
 &=& {\rm vec}(\a E)^T {\rm vec}( \a E\times \Sig^{-1} ) \\
&=& \v e^T ( \Sig_K\otimes \cdots \otimes \Sig_1)^{-1} \v e \ ,  \ \mbox{and} \\
| \Sig_K \otimes \cdots \otimes \Sig_1 |  &=&  
   \prod_{k=1}^K  |\Sig_k|^{n_k  } ,  
 \end{eqnarray*} 
where $n_k = \prod_{j:j\neq k} m_j$ is the 
number of columns of $\m Y_{(k)}$, i.e.\ the ``sample size'' for 
$\Sig_k$. 
$\Box$

{\sc Proof of Proposition 4.3.}  
We first obtain the full conditional distributions for the matrix normal case. 
Let $\a Y \sim {\rm anorm}(\a 0 , \Sig \otimes \Om ) $ and $\Sig^{-1}=\Ps$. 
Let $(\v a,\v b )$ form a partition of the row indices of $\a Y$, 
and  assume the rows of $\a Y$ are ordered according to this partition. 
The  quadratic term in the exponent of the density can then be written as 
\begin{eqnarray*}
\tr(\Om^{-1} \m Y^T  \Ps \m Y ) &=& 
  \tr \left ( \Om^{-1} ( \m Y_a^T \m Y_b^T )  
 \left (  \begin{array}{cc} \Ps_{aa} & \Ps_{ab} \\ \Ps_{ba} & \Ps_{bb} \end{array} \right )  \left ( \begin{array}{c} \m Y_a \\ \m Y_b\end{array} \right)  \right  )  \\ 
&=& \tr( \Om^{-1} \m Y_a^T \Ps_{aa} \m Y_a) + 2 \tr(\Om^{-1} \m Y_b^T \Ps_{ba} \m Y_a) +  \tr( \Om^{-1} \m Y_b^T \Ps_{bb} \m Y_b). 
\end{eqnarray*}
As a function of $\m Y_b$, this is equal to a constant plus the quadratic 
term of the matrix 
normal density with row and column covariance matrices of 
$\Ps_{bb}^{-1}$  and $\Om$ , and a mean of  
$ -\Om_{bb}^{-1} \Om_{ba} \m Y_a$.  
Standard results on inverses of partitioned matrices give 
the row variance as 
$\Ps_{bb}^{-1} = \Sig_{bb} - \Sig_{ba} (\Sig_{aa} )^{-1} \Sig_{ab}= \Sig_{b|a}$ and 
 the mean as $ -\Om_{bb}^{-1} \Om_{ba} \m Y_a = \Sig_{b|a} (\Sig_{aa})^{-1} \m Y_b$.   To obtain the result for the array case, note that 
if $\a Y \sim {\rm anorm}(\a 0, \Sig_1\circ \cdots \circ \Sig_K)$ then 
the distribution of $\a Y_{(1)}$ is matrix normal with 
row covariance $\Sig_1$ and column covariance 
$\Sig_K \otimes \cdots \otimes \Sig_2$. The conditional distribution can 
then be obtained by applying the result for the matrix normal case to 
$\a Y_{(1)}$ with $\Sig=\Sig_1$ and $\Om = \Sig_K \otimes \cdots \otimes \Sig_2$. 
$\Box$

\bibliographystyle{plainnat}
\bibliography{/Users/hoff/madrid/SharedFiles/refs}
\end{document}